\documentclass[twocolumn,showpacs,preprintnumbers,amsmath,amssymb,superscriptaddress]{revtex4}

\usepackage{graphicx}
\usepackage{dcolumn}
\usepackage{bm}
\usepackage{color}

\begin{document}


\title{Approach to Universality in Axisymmetric Bubble Pinch-Off}

\author{Stephan Gekle}
\author{Devaraj van der Meer}
\author{Jacco H. Snoeijer}
\author{Detlef Lohse}
\affiliation{
Department of Applied Physics and J.M. Burgers
Centre for Fluid Dynamics, University of Twente, P.O. Box 217,
7500 AE Enschede, The Netherlands
}

\date{\today}

\begin{abstract}
The pinch-off of an axisymmetric air bubble surrounded by an inviscid fluid is compared in four physical realizations: (i) cavity collapse in the wake of an impacting disc, (ii) gas bubbles injected through a small orifice, (iii) bubble rupture in a straining flow, and (iv) a bubble with an initially necked shape. Our boundary-integral simulations suggest that all systems eventually follow the universal behavior characterized by slowly varying exponents predicted in [Eggers \textit{et al.}, PRL \textbf{98}, 094502 (2007)]. However, the time scale for the onset of this final regime is found to vary by orders of magnitude depending on the system in question: while for the impacting disc it is well in the millisecond range, for the gas injection needle universal behavior sets in only a few microseconds before pinch-off. These findings reconcile the different views expressed in recent literature about the universal nature of bubble pinch-off. 
\end{abstract}


\maketitle


The precise nature of axisymmetric bubble collapse in a low-viscosity fluid has been a subject of controversy over the last years. Such a collapse may be initiated by a variety of different forces (e.g. surface tension, hydrostatic pressure, and external flows). In a later stage, however, it is only the requirement of mass conservation that forces the liquid to accelerate more and more as the shrinking bubble neck closes in on the axis of symmetry. This purely inertial nature of the final collapse motivated the first hypotheses about universality of the final collapse regime \cite{LonguetHigginsKermanLunde_JFM_1991, OguzProsperetti_JFM_1993}. A simple power-law was predicted with the neck radius scaling as the square root of the time remaining until the pinch-off singularity. Neither numerically nor experimentally could this behavior be confirmed. Instead, for different systems and initial conditions a variety of scaling exponents all slightly above 1/2 have been obtained \cite{GordilloEtAl_PRL_2005, BurtonWaldrepTaborek_PRL_2005, BergmannEtAl_PRL_2006,KeimEtAl_PRL_2006, ThoroddsenEtohTakehara_PhysFluids_2007a, GordilloSevillaMartinezBazan_PhysFluids_2007, BurtonTaborek_PRL_2008, Gordillo_PhysFluids_2008, BolanosJimenezEtAl_PhysFluids_2008} leading to doubts about the universal nature of bubble collapse.

Recently, the idea of universality has been revived by \cite{GordilloPerezSaborid_JFM_2006, EggersEtAl_PRL_2007} who suggested an intricate coupling between the radial and axial length scale. The authors of \cite{EggersEtAl_PRL_2007} explicitly predict the existence of a final universal regime which however is no longer a simple power-law, but characterized by a local exponent that slowly varies in time. The value of 1/2 is recovered in the asymptotic limit infinitesimally close to pinch-off. According to this theory the variety of observed exponents corresponds to different time averages of this local exponent.

In the present work we aim to reconcile the different views about universality in axisymmetric bubble pinch-off expressed over the last years. The key aspect is that we examine in detail how and when different physical realizations of bubble pinch-off reach the universal regime. We present detailed numerical simulations which are able to follow the neck evolution over more than 12 decades in time even for complex realistic systems. We first show that all systems that have recently been studied in the context of bubble pinch-off eventually follow the same universal behavior predicted by \cite{EggersEtAl_PRL_2007}. The time scale on which universality is reached, however, varies enormously from one setup to another. For an impacting disc \cite{BergmannEtAl_PRL_2006, BergmannEtAl_preprint} universality can be observed during several milliseconds prior to pinch-off and thus on a time scale which is  experimentally accessible. However, for gas bubbles injected through a small needle \cite{LonguetHigginsKermanLunde_JFM_1991, OguzProsperetti_JFM_1993, BurtonWaldrepTaborek_PRL_2005, KeimEtAl_PRL_2006, ThoroddsenEtohTakehara_PhysFluids_2007a, GordilloSevillaMartinezBazan_PhysFluids_2007, BurtonTaborek_PRL_2008, Gordillo_PhysFluids_2008, BolanosJimenezEtAl_PhysFluids_2008, SchmidtEtAl_preprint, TuritsynLaiZhang_preprint} universality sets in only a few microseconds (or even less, depending on the precise initial conditions) before pinch-off. This may well be the reason why universality has thus far never been observed even in very precise gas injection experiments and why non-inertial effects like surface tension have been claimed to play a dominant role in this geometry \cite{Gordillo_PhysFluids_2008, BolanosJimenezEtAl_PhysFluids_2008}.  

\begin{figure}
  \includegraphics{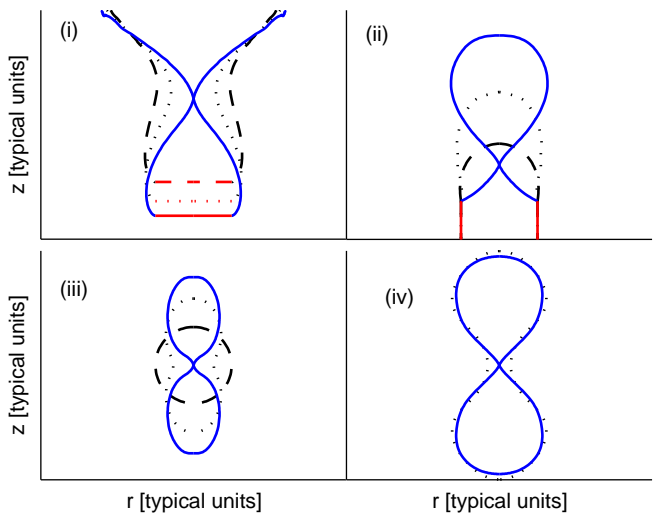}
  \caption{Illustration of the bubble collapse in the four different systems: impacting disc (i), gas injection through a needle orifice (ii), bubble in a straining flow (iii), and initially necked bubble (iv). Blue lines correspond to the free surface at pinch-off, while dashed and dotted black lines represent earlier bubble shapes. The disc and the needle are depicted in red.}\label{fig:illus}
\end{figure}

Four different physical systems have been reported in the literature on bubble pinch-off, numerically and experimentally, and will be compared in this Letter: 

(i) \textit{Impacting disc}. The bubble is created by the impact of a circular disc on a liquid surface \cite{BergmannEtAl_PRL_2006, BergmannEtAl_preprint} as shown in Fig.~\ref{fig:illus}(i). Upon impact an axisymmetric air cavity forms and eventually pinches off halfway down the cavity under the influence of hydrostatic pressure. Immediately after pinch-off the ejection of a violent jet can be observed whose formation however is not caused by the singularity alone \cite{GekleEtAl_PRL_2009} as one might naively expect. Since surface tension is negligible \cite{BergmannEtAl_PRL_2006, GekleEtAl_PRL_2009, BergmannEtAl_preprint} the only relevant control parameter is the Froude number $\mathrm{Fr}=V_0^2/gR_0$ with the impact velocity $V_0$, gravity $g$ and the disc radius $R_0$. In the data reported here the disc radius varies between 1 and 3cm and the impact velocity ranges from 1 to 20m/s. 

(ii) \textit{Gas injection through an orifice}. A small needle sticks through the bottom of a quiescent liquid pool \cite{LonguetHigginsKermanLunde_JFM_1991, OguzProsperetti_JFM_1993, BurtonWaldrepTaborek_PRL_2005, KeimEtAl_PRL_2006, ThoroddsenEtohTakehara_PhysFluids_2007a, GordilloSevillaMartinezBazan_PhysFluids_2007, BurtonTaborek_PRL_2008, Gordillo_PhysFluids_2008, BolanosJimenezEtAl_PhysFluids_2008, SchmidtEtAl_preprint, TuritsynLaiZhang_preprint} as illustrated in Fig.~\ref{fig:illus}(ii). A pressure reservoir connected to the needle slowly pushes a gas bubble out of the needle's orifice. The bubble then rises under the influence of buoyancy. When the air thread between the orifice and the main bubble becomes long enough, surface tension causes the thinning of the neck which eventually leads to the pinch-off of the bubble. We present data for three sample configurations A-C corresponding to Figs.~4,~10,~and~6 of \cite{OguzProsperetti_JFM_1993} and characterized by a Weber number $\mathrm{We_{A,B,C}}=$0.007, 36 and 173, respectively. (The Weber number is defined as $\mathrm{We}=\rho Q^2/(\pi^2a^3\sigma)$ with water density $\rho$, gas flow rate $Q$, needle radius $a$ and surface tension $\sigma$). 

(iii) \textit{Bubble in a straining flow}. The initially spherical bubble collapses due to a surrounding hyperbolic straining flow \cite{GordilloEtAl_PRL_2005, GordilloPerezSaborid_JFM_2006, RodriguezRodriguezGordilloMartinezBazan_JFM_2006, GordilloFontelos_PRL_2007}, see Fig.~\ref{fig:illus}(iii). 

(iv) \textit{Initially necked bubble}. Surface tension causes the pinch-off of a bubble starting off with an initially already pronounced neck \cite{EggersEtAl_PRL_2007} as illustrated in Fig.~\ref{fig:illus}(iv).

In all four systems we consider the idealized invisicid, axisymmetric bubble pinch-off neglecting the influence of the inner gas dynamics \cite{LeppinenLister_PhysFluids_2003, NitscheSteen_JComputPhys_2004, GordilloEtAl_PRL_2005, GordilloFontelos_PRL_2007, BurtonTaborek_PRL_2008, Gordillo_PhysFluids_2008, BolanosJimenezEtAl_PhysFluids_2008}, viscosity \cite{BurtonWaldrepTaborek_PRL_2005, ThoroddsenEtohTakehara_PhysFluids_2007a, Gordillo_PhysFluids_2008, BolanosJimenezEtAl_PhysFluids_2008}, and non-axisymmetric perturbations \cite{SchmidtEtAl_preprint, TuritsynLaiZhang_preprint}. For our numerical investigations we employ an axisymmetric boundary-integral code similar to the one described in \cite{BergmannEtAl_preprint} which has shown very good agreement with experiments of system (i) for various impact geometries \cite{BergmannEtAl_PRL_2006, GekleEtAl_PRL_2008, BergmannEtAl_preprint}. The validity of our implementation for the other systems is verified by comparison with the bubble shapes obtained in various earlier works \cite{OguzProsperetti_JFM_1993, RodriguezRodriguezGordilloMartinezBazan_JFM_2006, EggersEtAl_PRL_2007}. Some details about the simulation parameters are given in the EPAPS file \footnote{See EPAPS Document No. XXX for supplementary material. For more information on EPAPS, see http://www.aip.org/pubservs/epaps.html}. 

\begin{figure*}
  \includegraphics[width=1.9\columnwidth]{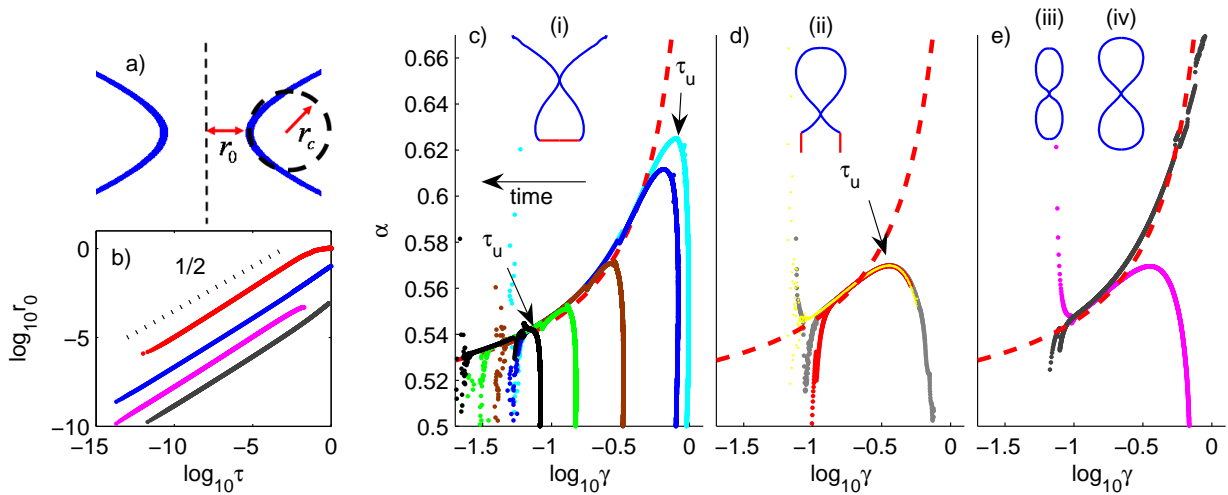}
  \caption{(a) Illustration of the cavity surface (blue line), the minimal neck radius $r_0$ and the local radius of curvature $r_c$. (b) ``Classical'' plot of the neck radius versus the time to pinch-off for system (i) with Fr=5.1 (blue), system (ii) in setup B (red), system (iii) (dark gray) and system (iv) (magenta). The curves are offset by one decade from each other to improve readability. (c) The local exponent $\alpha$ as a function of the aspect ratio $\gamma$ for system (i) with Fr=3.4 (cyan), Fr=5.1 (blue), Fr=46 (brown), Fr=500 (green), and Fr=4000 (black). After an initial transient all curves follow the same universal regime. The dashed line is Eq.~(\ref{eqn:alpha}). The local maxima correspond roughly to the start of the universal regime. (d) The local exponent for system (ii) in the three configurations: A (red), B (gray) and C (yellow). All curves A-C lie practically on top of each other. (e) The local exponent for system (iii) in dark gray and system (iv) in magenta follows the same universal behavior close to pinch-off. Small jumps in the data are due to the crossover between different node positioning algorithms employed in the initial and the final stages of the simulation (see EPAPS document [25]).}\label{fig:approach}
\end{figure*}

In a first approach to an analytical description of bubble collapse, the bubble shape can be approximated as an infinitely long cylinder (neglecting axial velocities) which yields a two-dimensional version of the well-known Rayleigh equation \cite{OguzProsperetti_JFM_1993, BurtonWaldrepTaborek_PRL_2005, GordilloEtAl_PRL_2005, BergmannEtAl_PRL_2006, GordilloPerezSaborid_JFM_2006} for the neck radius $r_0$
\begin{equation}
\frac{d\left(r_0\dot{r}_0\right)}{dt}\ln\frac{r_0}{R_\infty} + \frac{1}{2}\dot{r}_0^2 = F.
\label{eqn:Rayleigh}\end{equation}
Here, $F$ represents the force driving the collapse, and dots denote the derivative with respect to time $t$. $R_\infty$ is a cut-off radius required to saturate the pressure at large distances. Assuming $R_\infty$ to be a constant leads to the neck radius $r_0$  shrinking as a power-law (possibly with logarithmic corrections \cite{GordilloEtAl_PRL_2005, GordilloPerezSaborid_JFM_2006, BergmannEtAl_PRL_2006}) with exponent 1/2 as a function of the time to pinch-off $\tau=t_c-t$, where $t_c$ is the closure time. At first sight, this expectation seems to be very well confirmed for all four systems by the lines in Fig~\ref{fig:approach}~(b) which to the naked eye appear perfectly straight over more than 12 decades. The slope which corresponds to the scaling exponent is slightly larger than 1/2, in agreement with previous experiments and simulations \cite{GordilloEtAl_PRL_2005, BurtonWaldrepTaborek_PRL_2005, BergmannEtAl_PRL_2006, KeimEtAl_PRL_2006, GordilloPerezSaborid_JFM_2006,ThoroddsenEtohTakehara_PhysFluids_2007a, BurtonTaborek_PRL_2008, Gordillo_PhysFluids_2008, BolanosJimenezEtAl_PhysFluids_2008}.

A more detailed look at the local exponent, defined as the slope in Fig.~\ref{fig:approach} (b), $\alpha(\tau)=\partial \ln r_0 / \partial \ln\tau$ (Fig.~\ref{fig:approach} (c) -- (e)) however, reveals that the behavior of the neck radius cannot be described by a simple power-law. The local exponent $\alpha$ varies during the approach to pinch-off \cite{EggersEtAl_PRL_2007}. In fact, the relevant equation for the time-dependence of $\alpha$ in \cite{EggersEtAl_PRL_2007} can be derived directly from Eq.~(\ref{eqn:Rayleigh}) by letting $R_\infty=2\sqrt{r_0r_c}$. Here $r_c$ is the local axial radius of curvature (see Fig. 2(a)). The combination $\sqrt{r_0 r_c}$ is the scale by which the axial coordinate has to be rescaled in order to collapse neck profiles at different times when rescaling radial coordinates by $r_0$ \cite{BergmannEtAl_PRL_2006, ThoroddsenEtohTakehara_PhysFluids_2007a}. This leads to the aspect ratio of the cavity naturally being defined as $\gamma=r_0 / \sqrt{r_0r_c}$. With the above substitutions and working in the limit of vanishing $F$ we obtain from Eq.~(\ref{eqn:Rayleigh}):
\begin{equation}
\left(-\frac{d\alpha}{d\ln\tau} + \alpha - 2\alpha^2\right)\ln\left(\frac{4}{\gamma^2}\right) = -\alpha^2 \label{eqn:alpha}
\end{equation}
which is exactly identical to Eq.~(4) in \cite{EggersEtAl_PRL_2007} (being $\Gamma_1=8$ \cite{FontelosSnoeijerEggers_preprint} and $a_0''=2\gamma^2$ in the original notation). Equation (\ref{eqn:alpha}) with the $d\alpha/d\ln\tau$ term neglected due to the slow variation of $\alpha$ \footnote{In the numerical data one can verify that indeed $\left|d\alpha / d\ln\tau\right| \ll \left|\alpha-2\alpha^2\right|$.} is shown as the dashed red line in Figs.~\ref{fig:approach} (c)-(e). It represents the universal regime where the only driving is provided by inertia and all external forces have become negligible. We will now proceed to compare the approach of the different systems (i)-(iv) to this universal curve. For this we use the aspect ratio $\gamma$ as a universal ``clock'' replacing the time to pinch-off $\tau$ \cite{FontelosSnoeijerEggers_preprint}. Note that $\gamma\rightarrow 0$ as $\tau\rightarrow 0$ meaning that the cavity becomes more and more slender \cite{GordilloPerezSaborid_JFM_2006, ThoroddsenEtohTakehara_PhysFluids_2007a}.

One of the key points to address is if and how this behavior can be observed experimentally. Besides the obvious difficulty of obtaining a sufficient number of decades to observe the slow variation of the local exponent, the crucial question is: at what time (before pinch-off) does the system exhibit universal behavior? This is crucial because, firstly, the duration of the universal regime needs to be within the time resolution of the experimental equipment. And secondly, the onset of universality needs to happen before other effects such as air flow, viscosity, non-axisymmetric instabilities, etc.~unavoidably destroy the purely inertial regime. We will now provide those time scales for the various systems based on numerical boundary-integral simulations which do not have these limitations.

We start by considering the impacting disc system (i) in Fig.~\ref{fig:approach}~(c). It is evident that the data for all values of the control parameter follow -- after some initial transient -- the same universal curve in excellent agreement with Eq.~(\ref{eqn:alpha}). Figure~\ref{fig:approach}~(c) further gives us a good measure at what aspect ratio the universal regime is attained: approximately after passing their respective local maxima all curves follow the same behavior. The aspect ratio of this maximum can then easily be related to the physical time before pinch-off $\tau_u$. We find $\tau_u\approx 6$ms and $\tau_u\approx 1$ms for Fr=3.4 and Fr=4000, respectively. That the high Froude case reaches universality later can be intuitively understood: at high Froude the cavity closes deeper and therefore the hydrostatic driving pressure is larger and its effects on the neck dynamics can be felt longer. It is remarkable nevertheless that the duration of the universal regime changes only by a factor of less than 10 while the corresponding control parameter varies over three orders of magnitude. At the same time both values are easily within experimentally accessible time scales. 

We now compare this to system (ii), the bubble injection through a small needle \cite{LonguetHigginsKermanLunde_JFM_1991, OguzProsperetti_JFM_1993, BurtonWaldrepTaborek_PRL_2005, KeimEtAl_PRL_2006, ThoroddsenEtohTakehara_PhysFluids_2007a, GordilloSevillaMartinezBazan_PhysFluids_2007, BurtonTaborek_PRL_2008, Gordillo_PhysFluids_2008, BolanosJimenezEtAl_PhysFluids_2008, SchmidtEtAl_preprint, TuritsynLaiZhang_preprint} in Fig.~\ref{fig:approach}~(d). The approach to the universal regime is much less abrupt than in system (i). Due to this more gradual approach, it is difficult to specify precisely the time when universality is reached for the gas injection needle. We thus choose to keep our previous definition of $\tau_u$ being the time corresponding to the local maximum in Fig.~\ref{fig:approach}~(d). This gives a good upper bound for the time when universality sets in. Surprisingly, we find even these times to be of the order of 5$\mu$s in case A, 60ns in case C and as low as 10ns in case B, respectively
\footnote{The long duration for the quasi-static case A can be understood by considering that it has the smallest (non-dimensional) bubble size and thus the neck possesses the most symmetrical shape.}. 
Thus, the duration of the universal regime in the needle setup is dramatically (by at least three orders of magnitude) shorter than for the impacting disc. This may well explain why, among other experimental limitations, the universal regime has never been observed experimentally for this widely used system.

Figure~\ref{fig:approach} (e) confirms that also the systems (iii) and (iv) follow the universal regime. System (iii) does so even over the entire plotted range. Both are somewhat idealized systems and we are not aware of any experimental investigations regarding the approach to pinch-off in these systems. Without a relevant length and time scale it is therefore impossible to specify the physical time to universality in these cases. 

The different behaviors of the individual systems can intuitively be understood as follows. System (iii) contains no external driving force other than liquid inertia which makes it the ideal system to compare with Eq.~(\ref{eqn:alpha}). Indeed, this entirely inertial system follows the universal regime over the widest range in aspect ratios of all systems studied. Similarly, the very violent disc impact should introduce a large amount of inertia in system (i) which consequently follows the universal regime also for a rather long time. On the other hand, the two systems governed by surface tension (ii) and (iv) contain little inertia and thus approach the universal regime only relatively late and in a similar fashion.   

\begin{figure}
  \includegraphics[width=0.9\columnwidth]{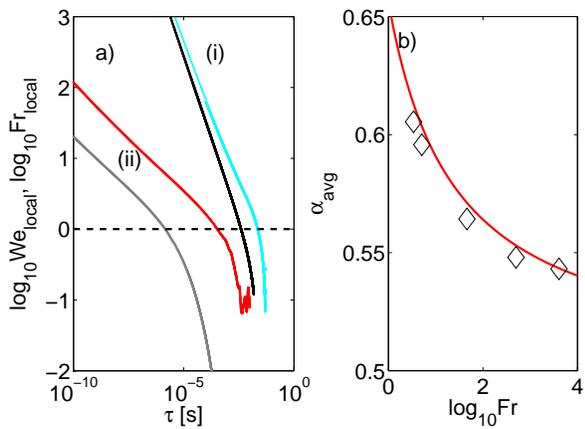}
  \caption{(a) The local Froude number for system (i) and an impact Froude number Fr=3.4 (cyan) and Fr=4000 (black) and the local Weber number for system (ii) with configuration A (red) and B (gray). The onset of the universal regime can be located roughly after the respective non-dimensional quantities have become larger than order unity (horizontal dashed line). (b) The averaged exponent measured in \cite{BergmannEtAl_PRL_2006} (black diamonds) is reproduced very well by our model with the constant $C=0.7$ (red line).}\label{fig:localNumbers}
\end{figure}

To make the above arguments more quantitative we realize that the universal regime sets in when the inertial driving of the collapse becomes dominant over the external driving force. This can be expressed by a local balance between inertia and the respective driving force. For system (i) the driving force is hydrostatic pressure and the relevant parameter thus the local Froude number $\mathrm{Fr_{local}}=\dot{r}_0^2/(gz_{c})$ with gravity $g$ and $z_c$ the depth below the surface where the cavity eventually closes. For system (ii) the local Weber number $\mathrm{We_{local}}=\rho \dot{r}_0^2 r_0/\sigma$ (with density $\rho$ and surface tension $\sigma$ of water) gives the balance between inertia and surface tension as the relevant driving force. The duration of the universal regime can then be estimated as the time before pinch-off when these local quantities become of order unity. Figure~\ref{fig:localNumbers} shows the local Froude and Weber numbers as a function of time to pinch-off $\tau$ for a number of representative cases of system (i) and (ii), respectively. One can clearly appreciate that $\mathrm{We_{local}}$ for the needle system becomes unity later than $\mathrm{Fr_{local}}$ for the impacting disc. This explains the large discrepancy in $\tau_u$ for the two systems. 

At the same time the distance between the two disc impacts with Fr=3.4 and Fr=4000 is much smaller than the distance between the two needle setups A and B. Accordingly, the duration of the universal regime varies only between $\sim$1ms and $\sim$6ms for the disc while its duration in the needle setup depends much stronger on initial conditions varying from micro- down to several nanoseconds as seen above. 

Our results allow us to explain the Froude-dependence of the experimentally and numerically observed exponent in \cite{BergmannEtAl_PRL_2006} for the impacting disc. Based on Fig.~\ref{fig:approach}~(c) these exponents can be viewed as a time-average of the local exponent. Due to the limited resolution and the onset of other effects (e.~g. air flow) only the right part of these plots is accessible in experiments and the time-average will be heavily weighted towards the beginning of the universal regime. We can thus assume the experimentally observed effective exponent to be rougly equal to the maximum value of the local exponent. Using Eq.~(\ref{eqn:alpha}) we can predict this value once the characteristic initial aspect ratio $\gamma_i$ for each cavity is known as a function of the impact Froude number.

This characteristic aspect ratio can be estimated to be proportional to the ratio between the (non-dimensional) maximum cavity expansion $R_{\mathrm{max}}$ and the closure depth $z_c$. The first quantity scales with the Froude number approximately as $R_\mathrm{max}\sim\mathrm{Fr}^{1/4}$ \cite{BergmannEtAl_preprint} while the latter behaves as $z_c\sim\mathrm{Fr}^{1/2}$ \cite{BergmannEtAl_preprint, DuclauxEtAl_JFM_2007}. The characteristic aspect ratio of the cavity thus becomes $\gamma_{i} = R_\mathrm{max}/z_c = C \mathrm{Fr}^{-1/4}$ with $C$ a constant of order unity. Inserting $\gamma_i$ into Eq.~(\ref{eqn:alpha}) and solving for $\alpha\left(\mathrm{Fr}\right)$, again neglecting $d\alpha/d\ln\tau$, we can thus predict the experimentally observable averaged exponent which is found in excellent agreement with \cite{BergmannEtAl_PRL_2006} as demonstrated by Fig.~\ref{fig:localNumbers} (b).

In conclusion, we have demonstrated that the universal theory of \cite{EggersEtAl_PRL_2007, FontelosSnoeijerEggers_preprint} faithfully predicts the approach of the neck radius for invisicid, axisymmetric bubble pinch-off in four different systems widely studied in the literature over the past years. Remarkably, however, the duration of the final regime is shown to be strongly dependent on the type of system and the various control parameters employed. While it lies easily within experimentally accessible time scales ($\sim$ms) for an impacting circular disc, it can be as low as a few nanoseconds for gas bubbles injected through a small orifice into a quiescent liquid pool. We were able to trace this difference back to the relative importance of the respective driving forces. Our findings reconcile the prediction of universality in bubble pinch-off \cite{GordilloPerezSaborid_JFM_2006, EggersEtAl_PRL_2007} with an apparent dependence on initial conditions \cite{BergmannEtAl_PRL_2006}, an apparently constant scaling exponent \cite{KeimEtAl_PRL_2006, ThoroddsenEtohTakehara_PhysFluids_2007a, BurtonTaborek_PRL_2008}, and with the observation that non-inertial forces can be dominant in many experimental settings \cite{Gordillo_PhysFluids_2008, BolanosJimenezEtAl_PhysFluids_2008, GordilloFontelos_PRL_2007}.

\begin{acknowledgments}
This work is part of the program of the Stichting FOM, which is financially supported by NWO.
\end{acknowledgments}

\end{document}